\documentclass[twocolumn,nobibnotes,nofootinbib]{revtex4}

\usepackage{amsmath,amssymb,bm}
\usepackage{epsfig}
\usepackage{graphicx}
\usepackage{pstricks,pst-node,pst-text,pst-3d}
\usepackage{subfigure}
\usepackage{hyperref}

\def\beq{\begin{equation}}
\def\eeq{\end{equation}}
\begin{document}

\title{Investigating the effects of the QCD dynamics in the neutrino absorption by the Earth's interior at ultrahigh  energies}
\author{V. P. Gon\c{c}alves$^{1,2}$ and D. R. Gratieri$^{3}$}

\affiliation{ $^{1}$ Department of Astronomy and Theoretical Physics, Lund University, 223-62 Lund, Sweden.}
\affiliation{$^{2}$ Instituto de F\'{\i}sica e Matem\'atica,  Universidade
Federal de Pelotas, 
Caixa Postal 354, CEP 96010-900, Pelotas, RS, Brazil}
\affiliation{ $^{3}$ Escola de Engenharia Industrial Metal\'urgica de Volta Redonda, Departamento de Ci\^encias Exatas,\\ Universidade Federal Fluminense, CEP 27255-125 - Volta Redonda, RJ - Brazil}


\begin{abstract}
The opacity of the Earth to incident ultra high energy neutrinos is directly connected with the behaviour of the neutrino - nucleon ($\sigma^{\nu N}$) cross sections in a kinematic range utterly  unexplored. In this work we investigate how the uncertainties in $\sigma^{\nu N}$ due  the different QCD dynamic models modify the neutrino absorption while they travel across the Earth.  In particular, we compare the predictions of two extreme scenarios  for the high energy behaviour of the cross section, which are consistent with the current experimental data. The first scenario considered is based on the solution of the linear DGLAP equations at small-$x$ and large-$Q^2$, while the second one  take into account the unitarity effects in the neutrino - nucleon cross section by the imposition of the Froissart bound behaviour in the nucleon structure functions at large energies. Our results indicate that probability of absorption and the angular distribution of neutrino events are  sensitive to the the QCD dynamics at ultra high energies. 
\end{abstract}

%
\maketitle
The observation of ultra  high energy (UHE) neutrino events at PeV by the IceCube Collaboration marks the birth of neutrino astronomy \cite{Ice-first, ice:science}. Astrophysical neutrinos are good messengers from sky. They have small cross-sections even at ultra high energies and hence they are weakly absorbed by the medium that  they travel. This property allows neutrinos to travel large distances trough the universe basically unperturbed, bringing to us information about the nature of the medium in which they are produced. Also neutrinos are not deflected by any magnetic field, and hence, when UHE neutrinos are detected in the Earth, the muon tracks produced into  detector  points to their source.   In this way, these astrophysical neutrinos would  help to solve the puzzles of what are the source of UHE particles as well as the production mechanism. In fact, the combination of  UHE neutrinos and cosmic rays in the so called multi-channel astroparticle analysis should allow to determinate the origin of such high energy particles.

 In order to interpret the experimental results is fundamental to take into account that the attenuation of the neutrino beam in route to a detector is strongly dependent on the high energy behaviour of the neutrino - nucleon cross section ($\sigma^{\nu N}$), which determines the opacity of  the Earth to incident neutrinos (For a review see, e.g. Ref. \cite{zeller}). As discussed by several authors in the last years
\cite{gqrs96,gqrs98,kma,pena,FJKPP,BBMT,kutak,jamal,magno,armesto,CCS,Victor-PRD83,Thorne,kniehl,kuroda,Victor-PRD88,Victor-PRD90,jeong,Albacete}, at ultra high energies, the neutrino-nucleon cross section  provides a probe of Quantum Chromodynamics (QCD) in the kinematic region of very small values of Bjorken-$x$ and large virtualities $Q^2$, which was not explored by the HERA measurements of the structure functions \cite{hera}. These studies demonstrated that the uncertainties present in the extrapolations of $\sigma^{\nu N}$  for this new kinematic range has direct impact in the event rate in high energy neutrino telescopes \cite{Victor-PRD83, Victor-PRD88, Albacete}.
In particular, the results from Ref. \cite{Victor-PRD83} shown that the solution of the linear Dokshitzer - Gribov - Lipatov - Altarelli - Parisi  (DGLAP) equation \cite{dglap} at small - $x$ and large $Q^2$ obtained in Ref. \cite{FJKPP}, denoted FJKPP hereafter,  provides an upper bound for the behaviour of  $\sigma^{\nu N}$ at ultra high energies. In contrast, the solution proposed in 
Ref. \cite{BBMT}, denoted BBMT hereafter, which imposes that $\sigma^{\nu N}$ satisfies the Froissart bound at high energies, can be considered a lower bound. As demonstrated in Ref. \cite{Victor-PRD83}, models that taken into account  the non - linear effects to the QCD dynamics predict high energy behaviours between these extreme scenarios.

 Our goal in this paper is to extend these previous studies  for the analysis of the probability of neutrino absorption by the Earth's interior at ultra high energies and determine the theoretical uncertainty present in this quantity. In particular, we compare our predictions with those obtained using the standard approach proposed in Refs.  \cite{gqrs96, gqrs98}, denoted GQRS hereafter.  For completeness, we also present the results for the absorption due to the Glashow resonance in the anti neutrino - electron scattering \cite{sheldon}. Our analysis is motivated by the fact that the 
 IceCube  \cite{ice:science}  and Antares \cite{antares} observatories are sensitive to neutrinos below  the horizon line. However, depending of the magnitude of the charged current neutrino interactions and the Glashow resonance, the Earth's can become fully  opaque to neutrinos with very high energies,  which implies that e.g. the  IceCube can becomes blind to neutrinos coming from north hemisphere \cite{soni}. Moreover, as neutrinos coming from different directions travels different distances and feels distinct matter potential,   the contribution of these interactions modify the attenuation effect, which should lead to distortion in the angular distributions of events. Both aspects motivate a detailed analysis of the neutrino absorption by the Earth.


Lets start our analysis presenting a brief review of the main formulas used to estimate the probability of neutrino absorption by the { Earth's} interior.
Following \cite{gqrs96} we define 
the probability of neutrino interaction while cross the Earth as
\begin{eqnarray}
P^{j}_{Shad}(E_{\nu})=\exp\left\{- \frac{ \,z_{j}(\theta_{z})}{{\cal{L}}^{j}_{int}}\right\}
\label{pshad1}
\end{eqnarray}
where $j = N, \, e$ and 
the interaction length for scattering with nucleons and electrons is given respectively by 
\begin{eqnarray}
{\cal{L}}^{N}_{int} &=&\frac{1}{N_{A}\sigma_{\nu N}(E_{\nu})} \nonumber \\
{\cal{L}}^{e}_{int} &=& \frac{1}{\langle Z/A \rangle N_{A}\sigma_{\bar \nu_{e}e}(E_{\nu})}~.
\end{eqnarray}
The amount of matter that neutrinos feels while travel across the Earth is  a function of the zenith angle $\theta_{z}$ 
\begin{equation}
z_{j}(\theta_{z})=\int^{r(\theta_z)}_{0}\rho_{j}(r)\,dr~,
\label{zfun}
\end{equation}
where $r(\theta_{z}) = -2 \,R_{Earth} \cos\theta_{z}$ is the total distance travelled by neutrinos,  $\rho_{j}(r)[g~cm^{-3}]$  is the density profile of the Earth.  In this work we use the density profile from \cite{PREM} and, following  \cite{giunti}, we define $N_{e}=N_{A}(\rho_{tot}/g)\langle Z/A \rangle$. The factor $\langle Z/A\rangle$ is the average ratio between electrons ($Z=e=p$) and nucleons  ($A=p+n$). We have that  $\langle Z/A \rangle=0.475$ for $r\le 3480$ km and  $\langle Z/A \rangle = 0.495$ for $r>3480$ km [See Fig. 10.26 from Ref.  \cite{giunti} for details].  In this way we can write
\begin{eqnarray}
\rho_{tot}(r) & = & \frac{N_{e}}{N_{A}}\frac{1}{\langle Z/A\rangle} \,[g/cm^{3}] \,\,, \nonumber \\
\rho_{e}(r) & = & \frac{N_{e}}{N_{A}}\,[g/cm^{3}] \,\,,
\end{eqnarray}
where $N_{A}=6.022 \times 10^{23}/mol=6.022 \times 10^{-23}$ CMWE (centimeters of water equivalent) is the Avogadro's number. For $\cos \theta_{z} = -1$ neutrinos crosses all the Earth, and Eq.~(\ref{zfun}) results in 10kt/cm$^{2}$, or $1\times10^{10}$ CMWE. Consequently, we can write 

\begin{eqnarray}
P^{j}_{Shad}(E_{\nu}) = \exp\left\{- \kappa_j ~\sigma_{\nu {j}}(E_{\nu})\int^{r (\theta_{z})}_{0}\rho_{j}(r)dr\right\} \,\,,
\end{eqnarray}
where $\kappa_N = N_{A}$ and $\kappa_e = \langle Z/A \rangle \cdot N_{A}$.
Finally, we can define the  absorption function for the neutrinos while it crosses the Earth as
\begin{widetext}
\begin{equation}
S^{j}(E_{\nu})=\int^{0}_{-1}~d\cos(\theta_{z})P^{j}_{shad}(E_{\nu})=\int^{0}_{-1}~d\cos(\theta_{z})~~ \exp\left\{- \kappa_j ~\sigma_{\nu {j}}(E_{\nu})\int^{r(\theta_{z})}_{0}\rho_{j}(r)dr\right\}\,\,.
\label{Senu}
\end{equation}
\end{widetext}

In what follows we will estimate $P^{j}_{shad}(E_{\nu})$ and $S^{j}(E_{\nu})$ considering different models for the (anti) neutrino - nucleon cross section and, for comparison, we also present the results for  (anti) neutrino - lepton interactions. Deep inelastic neutrino - nucleon scattering is described in terms of charged current (CC) and neutral current (NC) interactions, which proceed through $W^{\pm}$ and $Z^0$  exchanges, respectively.    As the neutral current (NC) interactions are sub-dominant,  we will consider in what follows, for simplicity, only charged current (CC) interactions. The total neutrino - nucleon  cross section is given by \cite{book}
\begin{eqnarray}
\sigma_{\nu N}^{CC} (E_\nu) = \int_{Q^2_{min}}^s dQ^2 \int_{Q^2/s}^{1} dx \frac{1}{x s} 
\frac{\partial^2 \sigma^{CC}}{\partial x \partial y}\,\,,
\label{total}
\end{eqnarray}
where $E_{\nu}$ is the neutrino energy, $s = 2 ME_{\nu}$ with $M$ the hadron mass, $y = Q^2/(xs)$ and $Q^2_{min}$ is the minimum value of $Q^2$ which is introduced in order to stay in the deep inelastic region. In what follows we assume $Q^2_{min} = 1$ GeV$^2$. Moreover, the differential cross section is expressed in terms of the nucleon structure functions $F_{i,CC}^N$ as follows \cite{book}
\begin{widetext}
\begin{eqnarray} 
\frac{\partial^2 \sigma_{\nu N}^{CC}}{\partial x \partial y} = \frac{G_F^2 M E_{\nu}}{\pi} \left(\frac{M_W^2}{M_W^2 + Q^2}\right)^2 \left[\frac{1+(1-y)^2}{2} \, F_{2,CC}^N(x,Q^2) - \frac{y^2}{2}F_{L,CC}^N(x,Q^2)+ y (1-\frac{y}{2})xF_{3,CC}^N(x,Q^2)\right]\,\,,
\label{difcross}
\end{eqnarray}
\end{widetext}
where $G_F$ is the Fermi constant and $M_W$ denotes the mass of the charged gauge boson. The calculation of $\sigma_{\nu h}$ involves  integrations over $x$ and $Q^2$, with the  integral being dominated by  the interaction with partons of lower $x$ and  $Q^2$ values of the order of the electroweak boson mass squared. 
In the QCD improved parton model the structure functions $F_2,  \,F_L$ and $F_3$ are calculated in terms of quark and gluon distribution functions. In this case the neutrino - nucleon cross section for charged current interactions on an isoscalar target is given by (See, e.g. Ref. \cite{book}):
\begin{widetext}
\begin{eqnarray} 
\frac{\partial^2 \sigma_{\nu N}^{CC}}{\partial x \partial y} = \frac{2 G_F^2 M E_{\nu}}{\pi} \left(\frac{M_W^2}{M_W^2 + Q^2}\right)^2 \left[
xq_N(x,Q^2) + x\bar{q}_N(x,Q^2)(1-y)^2\right]
\label{difcross_partonic}
\end{eqnarray}
\end{widetext}
with the quark and anti-quark densities given by $q_N = (d+u)/2 + s +b$ and $\bar{q}_N = (\bar{d}+\bar{u})/2 + c+ t$. 

\begin{figure}[t]
\includegraphics[scale=0.33,angle=270]{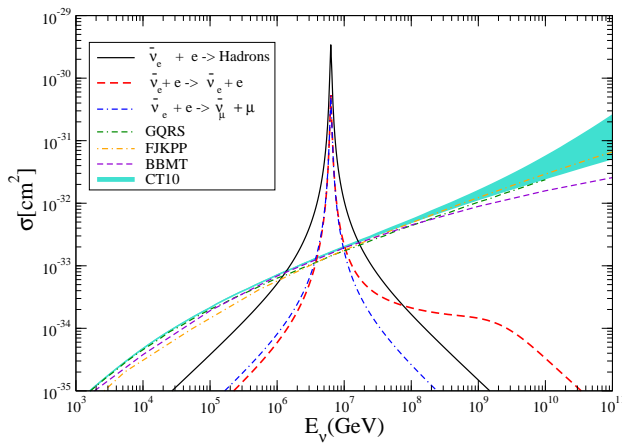}
\caption{(Color online) Comparison between the energy dependence predicted for the neutrino - nucleon cross-section  and for  the anti neutrino - electron  cross-section.}
\label{nucle-ress}
\end{figure}

The current estimates of the neutrino nucleon cross sections constrain the  structure functions and/or parton distributions using the HERA data and  are based on linear  QCD dynamics  (DGLAP or an unified DGLAP/BFKL evolution) \cite{gqrs96,gqrs98,kma,pena,FJKPP,Thorne}  or by models that taken into account the non-linear effects of QCD dynamics that area expected to be present at high energies \cite{BBMT,kutak,jamal,magno,armesto,CCS,Victor-PRD83,kniehl,Victor-PRD88,Victor-PRD90,Albacete}. In particular, the neutrino - nucleon cross section was originally calculated at leading order in Ref. \cite{gqrs96}, with the resulting parametrization being a benchmark for the evaluation of  UHE cosmic neutrinos. In Refs. \cite{CCS,Thorne} a next-to-leading order analysis was performed, and the uncertainties on high energy $\sigma_{\nu N}$ which are compatible with the conventional DGLAP formalism \cite{dglap} were estimated. Moreover, in Ref. \cite{FJKPP} it was estimated considering an analytical solution of the DGLAP equation, valid at  twist-2 and small-$x$, which implies  a power-like increasing of the cross section at ultra high energies. In contrast, in Ref. \cite{BBMT} the HERA data were successfully fitted assuming that the proton structure function saturates the Froissart bound, which implies $F_2^p \propto \ln^2 (1/x)$ and, consequently, that the increasing of $\sigma_{\nu N}^{CC}$ is smaller in comparison to the DGLAP predictions.
In Fig. \ref{nucle-ress} we  present a comparison between the predictions of the linear approaches (GQRS and FJKPP) and the Froissart-inspired model (BBMT) for the energy dependence of the neutrino nucleon CC cross section.  Moreover, for completeness of our analysis, we also present the predictions obtained the CT10 parametrization \cite{ct10} for the parton distributions (PDFs), derived using the DGLAP evolution equations, which allows to estimate the uncertainty present in the global fits as well as those associated to the extrapolation of the PDFs in a kinematical range beyond that probed by HERA, represented by the shaded band in the Fig. \ref{nucle-ress}. As expected from the solution of the DGLAP equation at small-$x$, the GQRS, FJKPP and CT10 models predict a strong increase of the cross section at ultra high energies, with the CT10 predictions being consistent with the GQRS and FJKPP results. Morever, the uncertainties present in the CT10 PDFs are fully propagated to the neutrino - nucleon cross section, with the size of the shaded band increasing at larger energies. 
Although these approaches agree at low energies, where the behaviour of the parton distributions are constrained by the HERA data, the GQRS and FJKPP differ by a factor 1.25 at $E_{\nu} = 10^{10}$ GeV and they are a factor 2 larger than the BBMT prediction. Moreover, at larger $E_{\nu}$, the FJKPP model predicts a strong increasing with the energy, differing from the BBMT prediction by a factor $\approx 3$ for $E_{\nu} = 10^{11}$ GeV. In comparison to the lower CT10 prediction, the BBMT one is smaller by a factor  $\approx 2$ for this neutrino energy. 
We have verified that the theoretical uncertainty increases for a factor $\approx 5.5$ (4.5) when we compare the FJKPP (CT10) and BBMT predictions for $E_{\nu} = 10^{13}$ GeV. 
It is important to emphasize that in Refs. \cite{FJKPP,BBMT} the authors have analysed the robustness of its results at large energies and estimated the uncertainties for the BBMT and FJKPP predictions  at $E_{\nu} = 10^{11}$ GeV as being smaller than 6\% and 14 \%, respectively.  We have that the resulting variations in the BBMT predictions are negligible comparable to the very large differences with respect to the FJKPP or CT10 predictions. 

In Fig. \ref{nucle-ress} we also present for comparison  the predictions for the anti neutrino - electron cross section, taking into account the presence of the Glashow resonance which is expected for neutrinos energies of the order of $ E_{\nu,res}=\frac{M^{2}_{W}}{2m_{e}}\approx 6.3 ~\mbox{PeV}$. Our predictions were obtained using the expressions for the cross sections presented in Refs. \cite{gqrs96,gqrs98,zeller} and the more recent values for the Weinberg angle, boson gauge masses and decay rates as given in Ref. \cite{PDG}. Our results demonstrate that anti neutrino - electron scattering  becomes equal or greater than CC neutrino-nucleon cross-section in the  energy range characterized by \,\,\,\, $10^6$ GeV $\le E_{\nu}\le 2 \times 10^7$ GeV.

\begin{figure}[t]
\includegraphics[scale=0.26,angle=270]{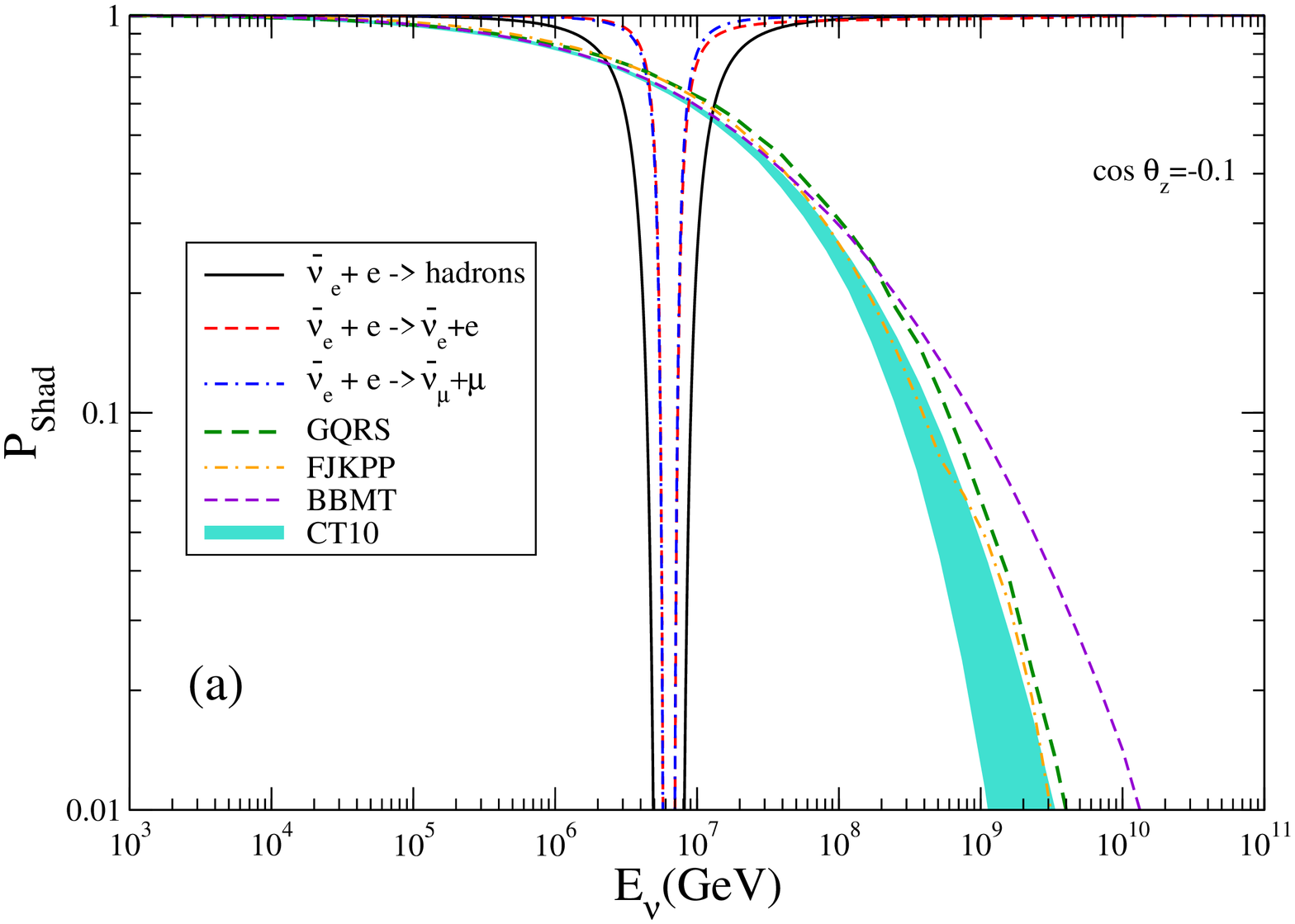} \\
\includegraphics[scale=0.26,angle=270]{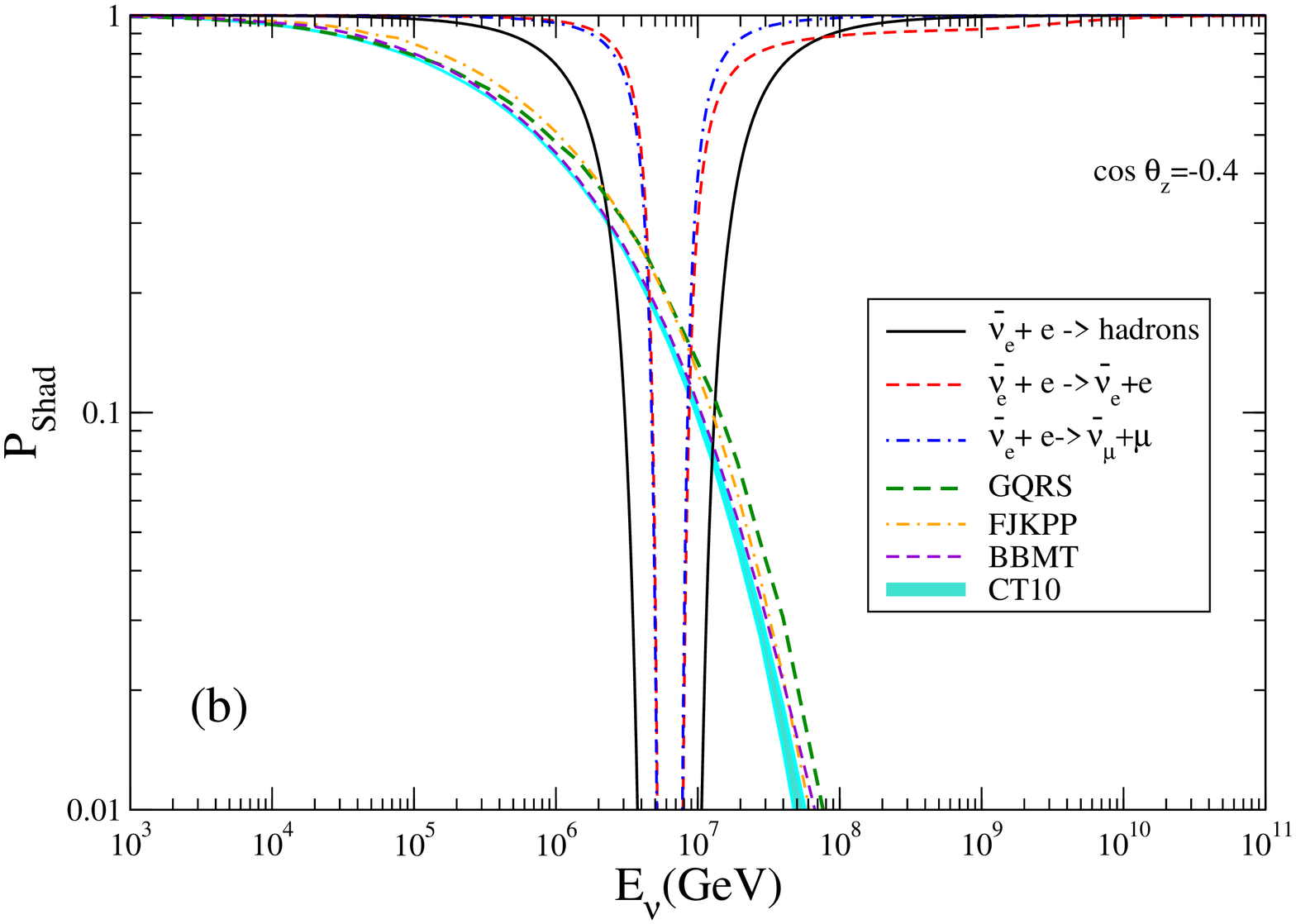} \\
\includegraphics[scale=0.26,angle=270]{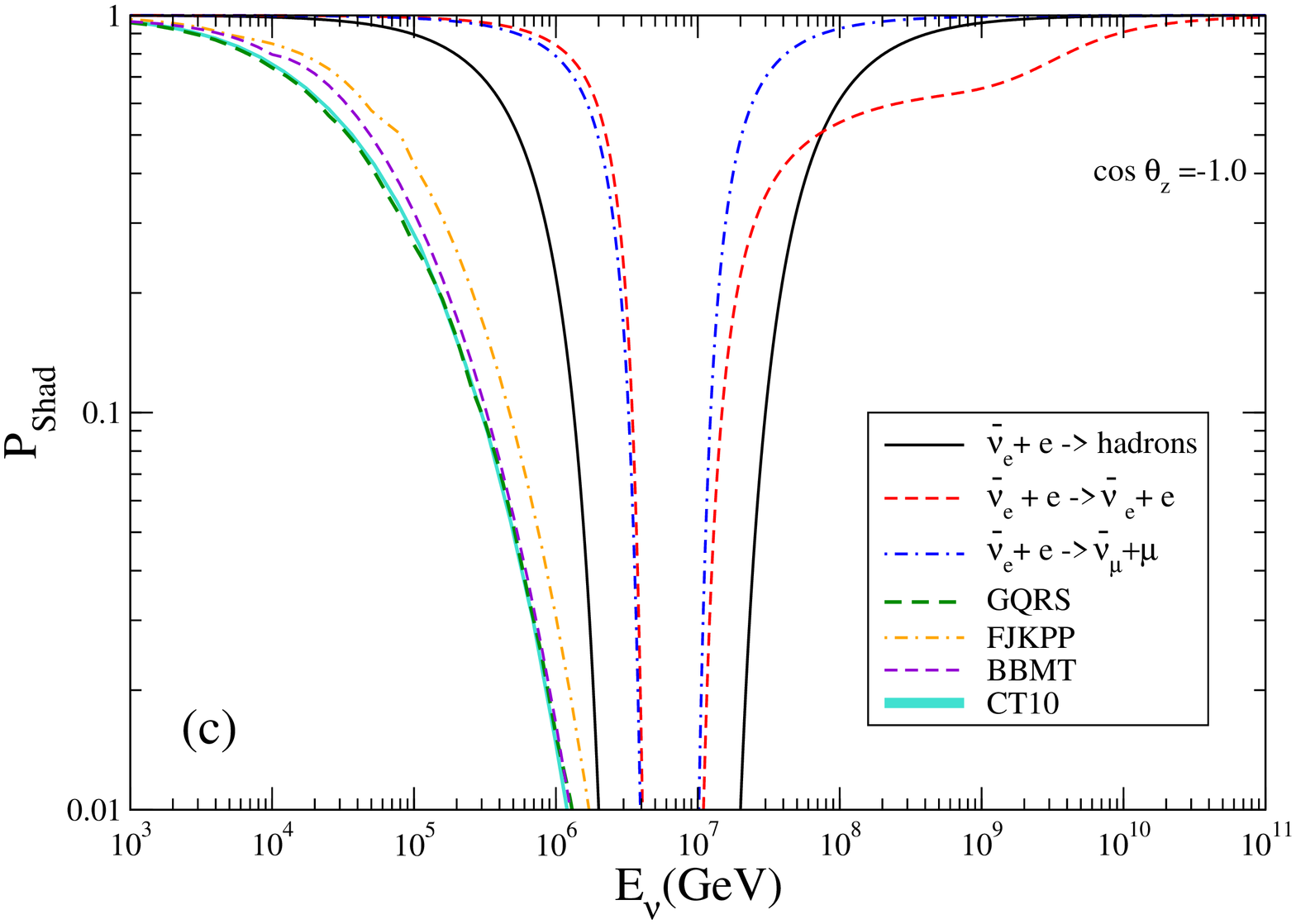}
\caption{(Color online) Energy dependence of the probability of  neutrino absorption for different values of the angle of neutrino incidence.}
\label{Pshad1}
\end{figure}

In Fig. \ref{Pshad1} present our results for the energy dependence of the probability of neutrino absorption considering different values of the 
zenith angle $\theta_{z}$. { We have found that} the peak of resonant $\bar{\nu}_e e$ cross section is translated into a maximum of absorption for  $E_{\nu} \approx 6.3 \times 10^{6}$ GeV, for all values of  $\cos \theta_{z}$. Basically, the angular effect in the resonant absorption is to enlarge the width of the resonance for $\cos \theta_{z}\rightarrow - 1$, { when} neutrinos travels greater distances inside Earth and experiment higher values of electron density. As expected, increasing the angle of the neutrino incidence, the higher density crossed by neutrinos amplify the effects of absorption due to CC neutrino-nucleon scattering, in such way that it dislocates the curves of $\nu N$ absorption to lower values of neutrino energy. We have that  the Earth becomes fully opaque to neutrinos for $\cos \theta_{z} = - 0.1$ is $E_{\nu}\approx 10^{10}$ GeV, while for  $\cos \theta_{z} = - 1.0$ it is $E_{\nu} \approx 10^{6}$ GeV. Consequently, the relative importance of neutrino - nucleon  absorption  to the Glashow resonance  depends of the angle of incidence of the neutrinos. Our results indicate that for $ \cos \theta_{z} = - 1.0$,  the attenuation due CC neutrino interactions becomes more important than the Glashow resonance even at IceCube energy range.  Moreover, the comparison between the distinct $\nu N$ predictions demonstrate that they can differ by $30\%$ ($55\%$) at $E_{\nu}=80$ ($300$) TeV, with  the Earth not being fully opaque to neutrinos  in this energy range even at $\cos \theta_{z} \rightarrow -1$, as indicated in  Fig. \ref{Pshad1} (c). On the other hand, for $ \cos \theta_{z} = - 0.1$ and ultra high energies, the distinct CC $\nu N$ predictions can be differ by  $\approx 100\%$. Basically, we obtain that the difference between these predictions is dependent on the zenith angle and the neutrino energy. Such uncertainty is not negligible and should be considered in the determination of the angular distribution of events in the IceCube and/or future observatories.

\begin{figure}[t]
\includegraphics[scale=0.33,angle=270]{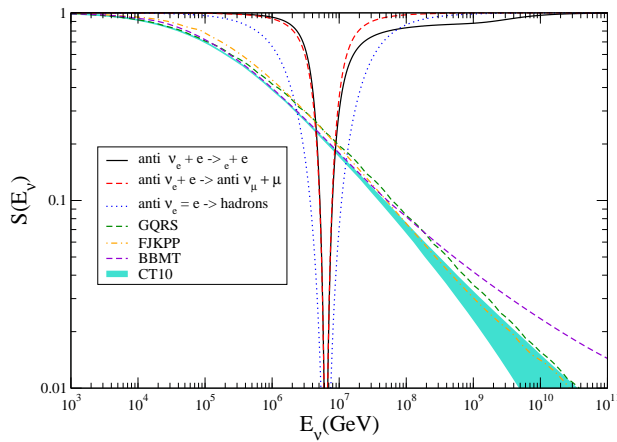}
\caption{(Color online) Energy dependence of the absorption function $S^j(E_{\nu})$.}
\label{fig:10}
\end{figure}


In Fig. \ref{fig:10}  we present our predictions for the 
absorption function $S^{j}(E_{\nu})$. We have that the integration over the  zenith angle tends to reduce  the energy range impacted by the  Glashow resonance absorption. Moreover, we have that the distinct predictions for $\nu N$ interactions are very similar for $E_{\nu}\le 10^{8}$ GeV, with the difference between the predictions reaching  $10\%$ at $80$ TeV.  On the other hand, at larger energies we have that the difference between the FJKPP (CT10) and BBMT predictions increases and becomes a factor 2 at $E_{\nu} \approx 10^{10}$ GeV, with the BBMT one being an upper bound. It is important to emphasize that considering the current estimates for the neutrino spectrum, which predict that the neutrino flux decreases with the energy with a power like behaviour, we have that the number of expected events at IceCube and/or future  observatories should be small for these energies. Therefore, the difference of a factor two between the predictions has a strong impact in the analysis and interpretation of the possible few events that should be observed.

Finally, lets summarize our main conclusions. In this paper we have estimated the impact of the current uncertainty in the description of $\nu N$ interactions at ultra high energies in the absorption of neutrinos crossing the  Earth until the detectors. Moreover, for comparison, the predictions considering $\bar{\nu}_e e$ were also presented. Our results indicated that the angular distribution of the neutrino events and the probability of absorption are sensitive to the treatment of the QCD dynamics at ultra high energies. Such results have direct implication in the determination of sources of UHE neutrinos below the horizon of IceCube neutrino observatory and in the analysis of the neutrino events in future observatories.

\section*{Acknowledgments} 
This work was  partially financed by the Brazilian funding
agencies CNPq, CAPES and FAPERGS.


\begin{thebibliography}{}

\bibitem{Ice-first}{M.G. Aartsen {\it et al.}  IceCube Collaboration. Phys. Rev. Lett. 111, 021103 (2013)}

\bibitem{ice:science} 
M.~G.~Aartsen {\it et al.} .IceCube Collaboration.
  Science {\bf 342}, no. 6161, 1242856 (2013).



\bibitem{zeller}
J. A. Formaggion and G. P. Zeller, Rev. Mod. Phys. {\bf 84}, 1307 (2012)

\bibitem{gqrs96}
R.~Gandhi, C.~Quigg, M.~H.~Reno and I.~Sarcevic,
  Astropart.\ Phys.\  {\bf 5}, 81 (1996)
  
\bibitem{gqrs98}
R.~Gandhi, C.~Quigg, M.~H.~Reno and I.~Sarcevic,
  Phys.\ Rev.\ D {\bf 58}, 093009 (1998)

  
  



\bibitem{kma}
  J.~Kwiecinski, A.~D.~Martin and A.~M.~Stasto,
  Phys.\ Rev.\  D {\bf 59}, 093002 (1999)


\bibitem{pena}
  J.~A.~Castro Pena, G.~Parente and E.~Zas,
  Phys.\ Lett.\  B {\bf 500}, 125 (2001);  Phys.\ Lett.\  B {\bf 507}, 231 (2001).

  
\bibitem{FJKPP}
  R.~Fiore, L.~L.~Jenkovszky, A.~Kotikov, F.~Paccanoni, A.~Papa and E.~Predazzi,
  Phys.\ Rev.\  D {\bf 68}, 093010 (2003); Phys.\ Rev.\  D {\bf 71}, 033002 (2005);  R.~Fiore, L.~L.~Jenkovszky, A.~V.~Kotikov, F.~Paccanoni and A.~Papa,
  Phys.\ Rev.\  D {\bf 73}, 053012 (2006). 
  
\bibitem{BBMT}
  E.~L.~Berger, M.~M.~Block, D.~W.~McKay and C.~I.~Tan,
  Phys.\ Rev.\  D {\bf 77}, 053007 (2008);
  M.~M.~Block, P.~Ha and D.~W.~McKay,
  Phys.\ Rev.\  D {\bf 82}, 077302 (2010); M.~M.~Block, L.~Durand, P.~Ha and D.~W.~McKay,
  Phys.\ Rev.\ D {\bf 88}, no. 1, 013003 (2013); M.~M.~Block, L.~Durand and P.~Ha,
  Phys.\ Rev.\ D {\bf 89}, no. 9, 094027 (2014)



\bibitem{kutak}
  K.~Kutak and J.~Kwiecinski,
  Eur.\ Phys.\ J.\  C {\bf 29}, 521 (2003)

\bibitem{jamal}
  J.~Jalilian-Marian,
  Phys.\ Rev.\  D {\bf 68}, 054005 (2003)
  [Erratum-ibid.\  D {\bf 70}, 079903 (2004)]; 
E.~M.~Henley and J.~Jalilian-Marian,
  Phys.\ Rev.\  D {\bf 73}, 094004 (2006)

\bibitem{magno}
  M.~V.~T.~Machado,
  Phys.\ Rev.\  D {\bf 70}, 053008 (2004); Phys.\ Rev.\  D {\bf 71}, 114009 (2005).
  
\bibitem{armesto}
  N.~Armesto, C.~Merino, G.~Parente and E.~Zas,
  Phys.\ Rev.\  D {\bf 77}, 013001 (2008)
  
\bibitem{CCS}  
 A.~Cooper-Sarkar and S.~Sarkar,
  JHEP {\bf 0801}, 075 (2008); A.~Cooper-Sarkar, P.~Mertsch and S.~Sarkar,
  JHEP {\bf 1108}, 042 (2011)
  
\bibitem{Victor-PRD83}
  V. P. Goncalves and P. Hepp, Phys. Rev. D  {\bf 83}, 014014 (2011).


\bibitem{Thorne}
 A. Connoly, R.S. Thorne and D. Watters, Phys. Rev D {\bf 83}, 113009 (2011).   

\bibitem{kniehl} 
  A.~Y.~Illarionov, B.~A.~Kniehl and A.~V.~Kotikov,
  Phys.\ Rev.\ Lett.\  {\bf 106}, 231802 (2011)

\bibitem{kuroda} 
  M.~Kuroda and D.~Schildknecht,
  Phys.\ Rev.\ D {\bf 88}, no. 5, 053007 (2013)

\bibitem{Victor-PRD88}
V.~P.~Goncalves and D.~R.~Gratieri,
  Phys.\ Rev.\ D {\bf 88}, no. 1, 014022 (2013)




\bibitem{Victor-PRD90}
V.~P.~Goncalves and D.~R.~Gratieri,
  Phys.\ Rev.\ D {\bf 90}, no. 5, 057502 (2014)

\bibitem{jeong} 
  Y.~S.~Jeong, C.~S.~Kim, M.~V.~Luu and M.~H.~Reno,
  JHEP {\bf 1411}, 025 (2014)

\bibitem{Albacete} 
  J.~L.~Albacete, J.~I.~Illana and A.~Soto-Ontoso,
  Phys.\ Rev.\ D {\bf 92}, no. 1, 014027 (2015)


\bibitem{hera}
  G.~Wolf,
  Rept.\ Prog.\ Phys.\  {\bf 73}, 116202 (2010)
  

\bibitem{dglap} V.N. Gribov and L.N. Lipatov, Sov. J. Nucl. Phys. {\bf 15}, 438 (1972);
G. Altarelli and G. Parisi, Nucl. Phys.  {\bf B126}, 298 (1977);
Yu.L. Dokshitzer, Sov. Phys. JETP {\bf 46}, 641 (1977).




\bibitem{sheldon} S.~L.~Glashow,
  Phys.\ Rev.\  {\bf 118}, 316 (1960).


\bibitem{antares}{S. Adrian-Mart\'ines {\it et al.} Antares Collaboration. Astrophys. Journ. Lett. {bf 786} L5, (2014)}

\bibitem{soni}{C. Chen, P. S. B. Dev. A. Soni. Phys. Rev D. {\bf 92}, 073001 (2014).}



\bibitem{PREM}{A. M. Dziewonski, D. L. Anderson, Physics of the Earth and
Planetary Interiors, {\bf 25}  297 (1981).}

\bibitem{giunti}{C. Giunti, C. W. Kim, ``Fundamentals of Neutrino Physics and Astrophysics'', Oxford University Press, (2007), Oxford.}

\bibitem{book}
  R.~Devenish and A.~Cooper-Sarkar,
  ``Deep inelastic scattering,''
Oxford University Press, (2004), Oxford. 


\bibitem{ct10} 
  H.~L.~Lai, M.~Guzzi, J.~Huston, Z.~Li, P.~M.~Nadolsky, J.~Pumplin and C.-P.~Yuan,
  Phys.\ Rev.\ D {\bf 82}, 074024 (2010)

\bibitem{PDG}
K.~A.~Olive {\it et al.} [Particle Data Group Collaboration],
  Chin.\ Phys.\ C {\bf 38}, 090001 (2014).


\end{thebibliography}
\end{document}